# CONSTRUCTION OF THE INITIAL PART OF A ION LINEAR ACCELERATOR FROM SIMILAR SHORT CAVITIES

V.V. Paramonov[†1], A.P. Durkin[1], L.Yu. Ovchinnikova[1,2], I.V. Rybakov[1]

[†1] - Institute of Nuclear Research RAS,
[2] - JSC "Research Institute "Ferrite-Domain"

# ПОСТРОЕНИЕ НАЧАЛЬНОЙ ЧАСТИ ЛИНЕЙНОГО УСКОРИТЕЛЯ ПУЧКОВ ИОНОВ ИЗ ОДНОТИПНЫХ КОРОТКИХ РЕЗОНАТОРОВ

В.В. Парамонов[†1], А.П. Дуркин[1], Л.Ю. Овчинникова[2,1], И.В. Рыбаков[1]

[1]-ФГБУН Институт Ядерных Исследований РАН,
[2]-АО «НИИ «Феррит-Домен»

*Abstract*

The construction of the initial part of a normally conducting linac for hydrogen ion beams with a pulsed current of ~ 20 mA up to an energy of ~ 70 MeV is considered. The RFQ at a frequency of ~ 160 MHz accelerates ions to an energy of ~ 4 MeV. Further acceleration is carried out at a doubled frequency by short, up to $5\beta\lambda$, cavities, operating in the TM010 mode, with drift tubes. Focusing is carried out by doublets of quadrupole lenses placed between the cavities. The structure of the accelerating-focusing channel, with given beam parameters, with reserves provides both the conditions for stable longitudinal and transverse motion of particles, and reliable technical implementation. The main results of the simulations of particle dynamics and the main parameters of the elements of the channel are presented. The possibility of constructing an linac with a higher output energy is analyzed.

*Аннотация*

Рассмотрено построение начальной часть нормально проводящего линейного ускорителя (ЛУ) пучков ионов водорода с импульсным током ~ 20 мА до энергии ~ 70 МэВ. Резонатор RFQ (Radio Frequency Quadrupole) на частоте ~ 160 МГц ускоряет ионы до энергии ~ 4 МэВ. Дальнейшее ускорение производится на удвоенной частоте короткими, длиной до $5\beta\lambda$, резонаторами на колебании TM010 с трубками дрейфа. Фокусировка осуществляется дублетами квадрупольных линз, размещенных между резонаторами. Структура ускоряюще – фокусирующего канала (УФК), при данных параметрах пучка, с запасами обеспечивает как условия устойчивого продольного и поперечного движений частиц, так и надежной технической реализации. Приводятся основные результаты расчета динамики частиц и основные параметры элементов УФК. Анализируется возможность построения ускорителя с более высокой выходной энергией.

## ВВЕДЕНИЕ

ЛУ пучков ионов водорода находят широкое применение как в качестве самостоятельных ЛУ класса мезонных фабрик, так и инжекторов в комплексы циклических ускорителей. Назначение ЛУ диктует требования к параметрам выходного пучка, что в свою очередь определяет решения по построению ЛУ.

В начальных частях ЛУ ионов водорода в области низких энергий традиционно используется структура Альвареца – резонаторы на колебании TM010 с трубками дрейфа, в которых размещены фокусирующие элементы (ФЭ) – квадрупольные линзы.

Для преодоления трудностей при сооружении, настройке структуры Альвареца для участка средних энергии предложена структура Separated Drift Tube Linac (SDTL), схема которой показана на Рис. 1, [1].

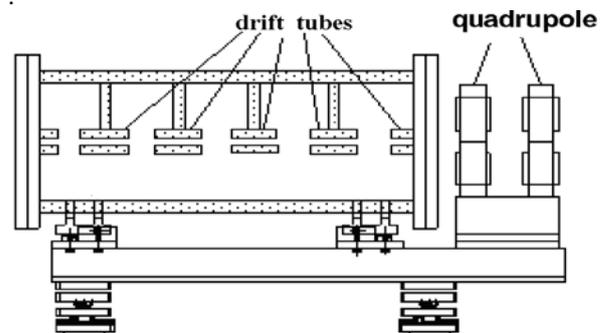

Рисунок 1: Схема резонатора структуры SDTL [1].

Резонаторы SDTL содержат небольшое число ускоряющих зазоров и ФЭ вынесены за пределы резонаторов. По ВЧ эффективности SDTL превосходит структуру Альвареца при энергии ионов >20 МэВ. В разработке ЛУ с вынесенными за пределы резонаторов ФЭ и выходной энергией > 30 МэВ практической альтернативы применению SDTL не известно. Достоинства SDTL стимулировали рассмотрения ее применения и на участках более низких энергий [2]. В данной работе обоснована возможность уверенного применения SDTL с энергии ионов от ~ 4 МэВ в ЛУ с умеренной величиной импульсного тока пучка.

## ПОСТРОЕНИЕ И ПАРАМЕТРЫ ЭЛЕМЕНТОВ УФК

В настоящее время неотъемлемой частью ЛУ ионов является RFQ, обеспечивающий захват непрерывного пучка из источника, группирование и ускорение сгустков до энергий в единицы МэВ. При равенстве рабочих частот RFQ и последующих резонаторов SDTL условия динамики сгустков существенно упрощаются. Но анализ ВЧ и технологических характеристик SDTL показывает преимущество ее применения на частотах выше 300 МГц. А в RFQ более выгодные условия динамики частиц и ВЧ характеристики обеспечиваются при более низких частотах.

Основные характеристики RFQ, выбранного из разработанных Коломийцем А.А. вариантов приведены в Таблице 1.

Таблица 1:Параметры RFQ.

| Параметр | Единицы | Значение |
|---|---|---|
| Частота | МГц | 162,5 |
| Входная энергия | кэВ | 60 |
| Напряжение | кВ | 140 |
| Напряженность E_smax/Ek |  | 1,8 |
| Синхронная фаза | Град. | -90, -25 |
| Прохождение пучка, I=15 мА | % | 99,41 |
| Захват пучка, I=15 мА | % | 99,0 |
| Поперечный аксептанс, норм | мм*мрад | 5,0 |
| Поперечный эмиттанс, норм., 6σ, I=15 мА | мм*мрад | 1,27 |
| Продольный эмиттанс, норм., rrms, I=15 мА | кэВ*град | 167,5 |
| Длина электродов | мм | 4741,2 |
| Коэффициент модуляции |  | 1,02-2,5 |
| Выходная энергия | МэВ | 3,996 |
| ВЧ мощность (расчет) | кВт/м | 81 |

При входном токе пучка 15 мА с нормализованном поперечным эмиттансом 1 π.мм.мрад резонатор с запасом обеспечивает условия для формирования и прохождения сгустков при умеренных длине и удельной мощности ВЧ потерь. Полученная без ухудшения других характеристик, величина продольного эмиттанса пучка влияет на дальнейшую разработку УФК ЛУ.

Следующий участок – линия согласования MEBT – длиной ~ 1680 мм содержит 5 квадрупольных линз с градиентами от 5,6 Т/ м до 28,9 Т/ м и двухзазорный резонатор SDTL на частоте 325 МГц в качестве группирователя. Длина линии MEBT достаточна для размещения дополнительного диагностического оборудования. На вход последующих резонаторов SDTL на частоте 325 МГц пучок поступает сжатым по фазе.

† paramono@inr.ru

Две последние линзы MEBT служат как входные линзы УФК SDTL.

Результаты работы [3] и собственные расчеты, см. Рис.2, показывают уверенную реализацию линз с интегральным градиентом до 5Т при уменьшенном до (25-35) мм диаметре апертуры и уменьшенной до 70 мм длиной полюсов. Для начала УФК такая величина интегрального градиента не требуется и применимы линзы с меньшей длиной полюсов.

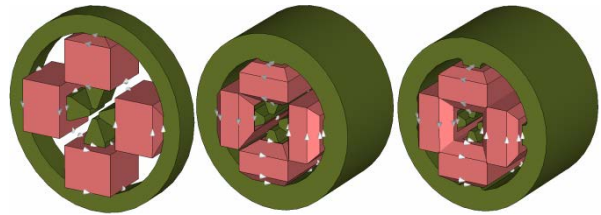

Рисунок 2: Модели электромагнитных квадрупольных линз, рассмотренных для УФК.

При реализуемых параметрах дублетов квадрупольных линз выбор схемы фокусировки ФДО дает меньшую, чем ФОДО, длину периода фокусировки и обеспечивает меньший радиус согласованного пучка. Наиболее критичным является начальный участок УФК. Для продольного движения частиц участок между резонаторами, на котором размещены дублеты ФЭ и технологические элементы, является участком дрейфа, приводящего к расплыванию сгустка по фазам. В начале УФК SDTL участки дрейфа занимают до половины длины периода фокусировки. По продольному движению периодическая повторяемость таких участков УФК приводит к сужению области устойчивого продольного движения частиц.

Анализ различных вариантов показал, что при энергии перехода 4 МэВ при разумной синхронной фазе захват в устойчивое дальнейшее ускорение возможен при наличии в резонаторе SDTL не более трех периодов ускорения. Синхронная фаза выбрана из анализа комплекса требований – обеспечение с запасом условий движения частиц при прогнозируемых параметрах входного пучка в УФК и учет технической реализуемости параметров оборудования для быстрейшего прохождения начального участка.

С ростом энергии частиц относительная длина резонатора в периоде фокусировки возрастает, а дефокусирующее действие ускоряющего поля снижается. Затухание фазовых колебаний с ростом энергии позволяет снижать по модулю значение синхронной фазы. Для сокращения длины ЛУ, с ростом энергии частиц осуществляется увеличение числа ускоряющих зазоров в резонаторах и повышение темпа ускорения.

Схематично резонаторы SDTL для различных участков УФК показаны на Рис. 3. В данной работе, по методике описанной в [2a], размеры элементов резонаторов оптимизированы для максимальной величины эффективного шунтового сопротивления Ze при ограничении Es/Ek <1.6, где Es –максимальная напряжен-ность электрического поля на поверхности, Ek =17,65 МВ – величина критерия Килпатрика на частоте 325 МГц. При Es/Ek >1.5 зависимость Ze(Es/Ek) выходит на плато и повышение Es/Ek >1.6, ведущее к повышению риска пробоев, не оправдывается повышением на единицы процентов величины Ze резонатора.

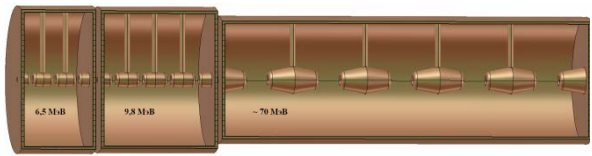

Рисунок 3: Резонаторы SDTL с различным числом ускоряющих зазоров для различных участков УФК.

Значения Es/Ek для резонаторов SDTL предлагаемого ЛУ показаны на графике Рис. 4а. На графике Рис. 4б показаны, рассчитанные с запасом в 25%, величины необходимой ВЧ мощности для резонаторов с учетом нагрузки пучком.

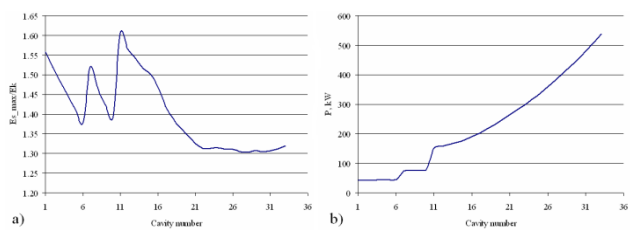

Рисунок 4: Величины Es/Ek (а) и необходимой ВЧ мощности для резонаторов SDTL .

Быстрый рост необходимой ВЧ мощности для последних резонаторов SDTL объясняется снижением погонного Ze структуры при энергии ионов >25 МэВ, [2], и выбранной для них достаточно высокой величины темпа ускорения ~ 2,7 МэВ/м.

Общие характеристики элементов УФК приведены в Таблице 2.

Таблица 2. Характеристики УФК и элементов.

| Параметр | Единицы | Значение |
|---|---|---|
| Начальная энергия | МэВ | 4 |
| Число резонаторов |  | 33 |
| Длины резонаторов отн. | βλ | 3,4,5 |
| Длины резонаторов | м | 0,26-1,7 |
| Синхронная фаза | градусы | -40, -30 |
| Темп ускорения EoT | МВ/м | 1,8-2,7 |
| Радиус апертуры | мм | 10, 12, 15 |
| Es/Ek |  | 1,6 – 1,3 |
| ВЧ мощность (+25%) | кВт | 42 - 500 |
| Добротность резонаторов |  | (43-55) 103 |
| Длины промежутков | βλ | 3 – 1,5 |
| Число дублетов |  | 32 |
| Интеграл. градиент линз | Т | 1,82-2,8 |
| Выходная энергия | МэВ | >70 |
| Длина УФК | м | <47 |

При энергии ионов свыше ~ 40 МэВ величина доступной ВЧ мощности является определяющей при выборе таких параметров как темп ускорения и число периодов в резонаторе. В данной работе мы опираемся на прогноз разработчиков мощных твердотельных ВЧ усилителей [4].

## СКВОЗНОЕ МОДЕЛИРОВАНИЕ ДИНАМИКИ ЧАСТИЦ

Сквозное моделирование динамики частиц в рассчитанных 3D распределениях электромагнитных полей резонаторов выполнено Коломийцем А.А. с помощью программы TRANSIT [5] от входа в RFQ до выхода SDTL для нулевого тока пучка и тока 15 мА.. Это позволило как подтвердить основные положения, заложенные в структуру УФК, так и наглядно увидеть влияние нелинейных эффектов, в частности пространственного заряда пучка.

На Рис. 5 для 99% частиц показаны огибающие пучка в УФК SDTL при токе 15 мА.

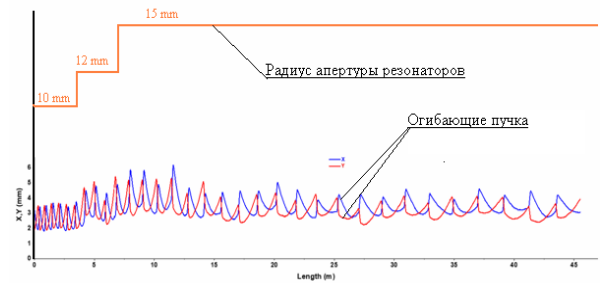

Рисунок 5: Огибающие пучка в УФК SDTL для 99% частиц при токе пучка 15 мА.

Проходя через RFQ и линию MEBT, на вход SDTL пучок поступает с шириной по фазам ~ 40 градусов с учетом существенных хвостов распределения. Это позволяет осуществить полный захват пучка для ускорения в SDTL с начальной синхронной фазой -40 градусов. В ускорении до энергии ~ 10 МэВ пучок имеет практически начальную ширину по фазам, что объясняется значительной длиной промежутков между резонаторами на данном участке УФК. Частицы, находящиеся на краях входного распределения по фазам, движутся по близким к сепаратрисе траекториям и формируют ореол продольного эмиттанса.

При дальнейшем ускорении влияние дрейфовых промежутков понижается и фазовая ширина сгустков уменьшается. Это позволяет изменить значение синхронной фазы до более выгодных величин.

Влияние сил пространственного заряда наиболее существенно в начале УФК. Происходит как снижение, так и разброс частот поперечных колебаний частиц. Это выражается как в росте огибающих пучка, так и естественном увеличении поперечного эмиттанса. Поэтому предусмотрено увеличение радиусов апертуры резонаторов, см. Рис. 5 и ФЭ на различных участках УФК. Обеспечивается необходимый значительных запас по отношению аксептанса УФК к ожидаемому эмиттансу пучка. В результаты численного моделирования динамики пучка потерь частиц от выхода RFQ до выхода SDTL не обнаружено.

Фазовые портреты выходного пучка из УФК SDTL при токе 15 мА показаны на Рис. 6.

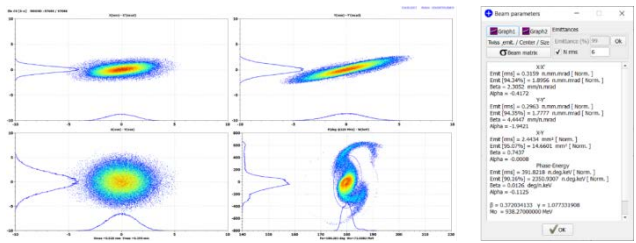

Рисунок 6: Фазовые портреты пучка на выходе УФК SDTL при токе 15 мА.

Результаты расчетов показывают, что структура SDTL может быть успешно использована для ускорения ионов водорода непосредственно после RFQ с выходной энергией 4 МэВ и небольшим продольным эмиттансом пучка. Структура УФК, при данных параметрах пучка, с запасами обеспечивает как условия устойчивого продольного и поперечного движений частиц при незначительном, 43% от выхода RFQ, росте эмиттанса пучка.

## ОБСУЖДЕНИЕ.

Приведенные результаты являются физическим обоснованием гарантированной возможности применения SDTL непосредственно после RFQ с выходной энергией 4 МЭВ при умеренных величинах тока и продольного эмиттанса пучка. Могут быть предложены, в том числе и авторами данной работы, и другие решения, как по составу оборудования, так и границам между участками рассматриваемого УФК. Основное внимание уделялось сбалансированности предлагаемых решений как по динамике частиц, так и параметрам оборудования, при гарантированных возможностях обеспечения характеристик пучка и изготовления оборудования.

Естественно, в случае разработки технического предложения необходима дальнейшая взаимосвязанная оптимизация как динамики частиц, так и параметров оборудования с целью купирования просматриваемых 'узких мест' и унификации параметров резонаторов.

Для более высоких энергий ионов SDTL, частота 324 МГц, применена, [6], в ЛУ комплекса J-PARC на участке энергий от 50 МэВ до 191.5 МэВ. В модификации связанных резонаторов (CSDTL) структура на частоте 325 МГц применена на участке от 50 МэВ до 104 МэВ в ЛУ Linac 4 CERN [7]. ЛУ J-PARC и ЛУ Linac 4 CERN разработаны для более высоких величин тока пучка. Это доказывает применимость SDTL до энергий ионов ~ (190 – 200) МэВ. Причем при сооружении уже действующих упомянутых ЛУ накоплен как зарубежный, [6], так и отечественный, [7], опыт изготовления резонаторов.

ВЧ эффективность SDTL, т.е. величина Ze, после максимума при энергии ионов ~ 20 МэВ постепенно снижается. В области энергий выше (100 – 150) МэВ SDTL проигрывает по ВЧ эффективности структурам на основе связанных ячеек.

Набор применяемых в ЛУ структур определяется необходимой выходной энергией частиц.

При выходной энергии ЛУ равной 200 МэВ, с точки зрения максимальной ВЧ эффективности структур, речь идет о замене всего нескольких резонаторов SDTL на всего несколько резонаторов совершенно отличной ускоряющей структуры. Выигрыш в ВЧ эффективности относительно незначительный и экономически не компенсирует дополнительные затраты на исследование, разработку, освоение изготовления технологически отличающейся структуры

Но если необходимая выходная энергия ЛУ равна 300 МэВ или выше, переход на другую ускоряющую структуру, например [8], экономически оправдан.

## ЗАКЛЮЧЕНИЕ

Ускоряющая структура SDTL предложена для участка 'средних' энергий ЛУ ионов водорода. В разработке ЛУ с энергией свыше 50 МэВ, при условии вынесенных за пределы резонаторов фокусирующих элементов, разумной альтернативы применению SDTL не известно. Применение SDTL до энергий ~200 МэВ доказано практикой действующего ЛУ J-PARC. Для умеренной величины тока пучка в работе обоснована применимость SDTL непосредственно после RFQ с энергий 4 МэВ. С запасами обеспечиваются необходимые условия динамики частиц при надежной технической реализации оборудования.

Технико-экономические преимущества применения в ЛУ единой ускоряющей структуры, включая техническую разработку, технологическое освоение, изготовление резонаторов, монтаж в ЛУ, настройку, запуск и дальнейшую эксплуатацию, неоспоримы.